\documentclass[aps,pre,groupedaddress,superscriptaddress,showpacs,showkeys,preprintnumbers,preprint]{revtex4}

\usepackage{graphicx}

\begin{document}
 \preprint{GCEP-CFUM2004/AC03}

 \title{KINETICS OF DIFFUSION-LIMITED REACTIONS WITH BIASED DIFFUSION IN PERCOLATING TO COMPACT SUBSTRATES}

 \author{N.J. Gon\c{c}alves}
  \affiliation{Departamento de F\'{i}sica, Universidade de Tr\'{a}s-os-Montes e Alto Douro, Apartado 1013, 5000-261 Vila Real, Portugal}
  \affiliation{GCEP - Centro de F\'{i}sica da Universidade do Minho, 4710-057 Braga, Portugal}
 \author{J.A.M.S. Duarte}
  \affiliation{Centro de Matem\'{a}tica da Universidade do Porto, 4169-007 Porto, Portugal}
 \author{A. Cadilhe}
  \email[]{cadilhe@fisica.uminho.pt}
  \affiliation{GCEP - Centro de F\'{i}sica da Universidade do Minho, 4710-057 Braga, Portugal}

 \date{\today}

 \begin{abstract}
  We studied through Monte Carlo simulations, the kinetics of the two-species diffusion-limited reaction model with same species excluded volume interaction in substrates embedded on a square lattice ranging in occupancy from a fractal percolating structure to the compact limit.
  We study the time evolution of the concentration of single-particle species for various values of substrate occupancies, $0.5927460$, $0.61$, $0.63$, $0.65$, $0.7$, $0.8$, and $1$, where the first value corresponds to the percolating probability of the square lattice.
  We show that in the diffusion-limited reaction regime, the kinetics strongly depends on the presence of a bias along a particular square lattice direction, representing the net effect of a driving field.
  We were able to explain the slow dynamics at high values of the driving field in terms of \emph{traps} appearing in diluted substrates, particularly at the percolation threshold.
 \end{abstract}



 \maketitle

 \section{Introduction}
  Non-equilibrium phenomena are ubiquitous in nature and of great interest to model \cite{Marro99,Bunde91,benAvraham00,Privman94,Privman96}.
  Such a model of non-equilibrium phenomena is the so-called diffusion-limited reaction model \cite{Ovchinnikov72,Torney83:a,Toussaint83,Kang84:a,Kang84:b,Kuzovkov88,Privman96,Cadilhe97,benAvraham00}.
  This model applies, when the time for two particles to react is less than the time for the particles to diffuse between collision.
  In this limit, diffusion becomes the slowest, rate determining process.
  There has been a host of results in the last two decades regarding the various properties of these reactions, and the realm of application exceeds what the name suggests.
  In fact, diffusion-limited reactions, apart from an interest on their own as case study of non-equilibrium phenomena, have been proved useful in the understanding of a wide range of natural phenomena like colloids and aerosols dynamics, soliton-antisoliton dynamics in quasi-$1d$ systems, vapor-deposited thin films, exciton reactions, and electron-hole recombination in semiconductors\cite{benAvraham90,Friedlander77,Family89,Kopelman87,Kopelman88,Schiff95}.
  The underlying reason for theoretical interest is related to the failure of the simple minded classical rate equation approach to the study of diffusion-limited phenomena.
  More specifically, the kinetics of these reactions is dominated by fluctuations in particle density, which cannot be grasped by a mean-field, rate-equation type of approach.
  For example, the bimolecular reaction $A+B\rightarrow\emptyset$ has an upper critical spatial dimension $d_c=4$, which means that below four spatial dimensions, the kinetics is dominated by fluctuations in particle density.
  As fluctuations are more effective on lower spatial dimensions, there are a host of results in $d=1$ \cite{Torney83:b,Racz85,Bramson88,benAvraham90,Privman92,Privman96,Cadilhe97}, but some research has also been done in two and three dimensions \cite{Toussaint83,Kang84:a,Kang84:b,Bramson88, benAvraham88,Kuzovkov88,benAvraham90}.
  In the present work, we focus our attention to the relevant cases of the effect of biased diffusion \cite{Kang84:a,Janowsky95:a,Janowsky95:b,Ispolatov95,Cadilhe97,Shafrir01}, as a consequence of a driving field, on the kinetics of diffusion-limited reactions on substrates ranging from the diluted, percolating cluster \cite{Kang84:a,Meakin84,Anacker87,Argyrakis92} to the compact substrate.
  We perform detailed analysis of the various stages of evolution of the kinetics in the isotropic and biased diffusion cases, and for different substrate occupancies.
  We show novel interesting properties of these models including trap-escape dominated kinetics.

  The paper is organized as follows: In Section~\ref{modelsimulations}, we define the model and present the details of our simulations.
  Next, in Section~\ref{resultsdiscussion}, we present our preliminary results and discuss them.
  Finally, in Section~\ref{conclusion}, we present our conclusions.

 \newpage
 \section{Model and Simulations}\label{modelsimulations}
  To simulate the model, we start by defining it in a square lattice with periodic boundary conditions for concreteness, but generalization to other lattices is straightforward.
  According to the percolation model each lattice site is occupied with probability $p$ and left empty with probability $1-p$ \cite{Stauffer79,Stauffer94}.
  For values of the occupation probability $p\geq p_c$, where $p_c$ is the percolation threshold probability, the largest cluster thus formed by nearest-neighbor sites has, at least, one path connecting opposite borders.
  In this percolative cluster, we deposit a uniform initial concentration of $A$- and $B$-type of reactive species, $c_0$, with equal number of both particle species.

  The kinetics of the system takes place on the percolating cluster defined above.
  Evolution takes place (Fig.~\ref{model}~a)) by randomly choosing a particle and allowing it to attempt to move to one of its nearest-neighbor sites.
  At every hopping attempt, only four events can take place.
  For example, a particle can diffuse by hopping to a nearest neighbor unoccupied site belonging to the cluster.
  As a result, in a diffusive move a particle solely changes its position (Fig.~\ref{model}~a)-i)).
  However, if the neighboring site is occupied by a like-species particle, excluded volume interaction prevents the particle from moving and the attempt fails, leaving both particles in their original positions (Fig.~\ref{model}~a)-ii)).
  A failed hopping attempt also takes place when the chosen nearest neighboring site belongs to the perimeter of the percolating cluster (Fig.~\ref{model}~a)-iii)), and, as a result, the particle just remains in its original position.
  This is equivalent to having reflective boundary conditions within the percolating cluster.
  The last event occurs when the selected site hops onto an unlike-species particle.
  In this case, both particles react and desorb from the substrate (Fig.~\ref{model}~a)-iv)), i.e., $A+B\rightarrow\emptyset$.
  Finally, time is incremented, by the reciprocal of the remaining number of particles, whether an attempt is successful.

  In this work we simulated systems, all embedded in a square lattice with a size of $1024^2$ sites, with a substrate occupancy ranging from the percolation threshold given by $p_c=0.5927460$ up to the compact case with $p = 1$.
  The intermediate values of the substrate occupancy, $p$, we simulated are $0.61$, $0.63$, $0.65$, $0.70$, and $0.80$.
  Applying an external field along one of the square lattice directions, causes the number of attempts of a particle to increase along this direction as compared to attempts in the opposite direction.
  The direction of the field is kept fixed along a square lattice direction, so, every $n$ attempt in the field direction, a particle attempts to move, on average, $p_f n$ times along the field direction and $p_b n$ times in the opposite direction, where $p_f$ and $p_b$ are the probabilities of a particle to move, respectively, forward and backwards along the field direction ($p_f + p_b= 1$).
  We further define $\eta= p_f - p_b$, which describes the bias introduced by the presence of the field.
  As defined, the bias $\eta= 0$, corresponds to zero field condition, i.e., isotropic diffusion, while for $\eta= 1$ we have a full bias condition.
  Perpendicularly to the field direction, diffusion is isotropic.
  Therefore, the probability, $v_i$, for a particle to attempt to move along any of the four nearest-neighbor directions of the square lattice is given by
   \begin{equation}\label{HopDist}
    v_{i} =
    \left\{
     \begin{array}{l@{\quad\quad}l}
      \frac{1+\eta}{4}&\mbox{along}\\
      \frac{1-\eta}{4}&\mbox{opposite to}\\
      \frac{1}{4}&\mbox{perpendicular to} \\
     \end{array}\\\quad
    \right.
    \mbox{the field direction.}
   \end{equation}
  In the present work, the values of $\eta$ used in simulations are $0$, $0.4$, $0.8$, and $0.9$.

  The direct consequence of the presence of bias, is to favor the rapid build up of correlations along the direction of the field as compared, for example, to a perpendicular direction.
  Consequently, it is appropriate to use rectangularly shaped lattices, which take into account the above fact with the longer side of the substrate along the field, as expected: we take, $L_\parallel/L_\perp= (1+\eta)$, where $L_\parallel$ and $L_\perp$ correspond to the lattice linear dimension along the field and perpendicular to it, respectively.

  A set of five samples were simulated for each pair of parameters, $(p, \eta)$, with an initial total concentration, $c_0= 0.4$, with periodic boundary conditions.

 \newpage
 \section{Results and Discussion}\label{resultsdiscussion}
  We present our results for the time evolution of single-species concentration in two limiting cases, namely zero, $\eta= 0$ and strong ,$\eta = 0.95$, driving fields for all the $p$ values presented above.
  We start by presenting the time dependence of the concentration, as shown in Fig.~\ref{c_zero}, in the isotropic, zero field case.
  We observe three major distinct evolution stages (Fig.~\ref{c_zero}~-~inset).
  The first stage lasts until the concentration drops by $18\%$ at $p_c$ and by $27\%$ in the compact system, and kinetics is well-characterized by rate equations.
  This is so, since the distribution of both particle species is homogeneous, thus spatial fluctuations are unimportant.
  Support for this line of reasoning comes from the fact that curves from different substrate occupancies differ only slightly within this stage, a clear indication of the lack of sensitivity on the details of the substrate.
  Moreover, for every simulation we also adopted a different substrate, so evolution is also insensitive to the substrate structure.
  Being on the first state decaying, the kinetics is now dominated by fluctuations on particle density, i.e., the diffusion-limited regime.
  As particles are depleted, i.e., annihilated during the first stage, domains start to form due to fluctuations in particle number.
  Regions rich in $A$-particles form a domain of such particles, while domains of $B$-particles appear in regions with deficiency of $A$-particles.
  Moreover, particles have to diffuse until they collide, which represents the slowest, rate determining, process present in the system.
  The kinetics of these systems is profoundly related to the initial conditions, both spatial and temporal correlations grow over time due to local, initial fluctuations in particle-number of each species.
  The concentration decays as a power-law, $t^{-\alpha}$, where the exponent $\alpha$ takes a value of $0.5$, for the compact system in 2-D.
  In this regime, mean-field, classical rate equations do $\underline{\mbox{not}}$ describe the kinetics of the system well, since from these a concentration decay with $\alpha=1$ is obtained.
  Rate equations are not able to capture the spatial fluctuations in particle number, but these are fundamental for the proper characterization of the kinetics of diffusion-limited reaction systems.
  At the percolation threshold $\alpha\simeq 0.347$, which agrees with the result quoted in reference \cite{Argyrakis92,Goncalves04:b}.
  The late times stage corresponds to a concentration lower than $10^{-3}$, it is a consequence of the finiteness of the system, and it decays exponentially.

  The various stages can be identified, including the value of $\alpha$, by plotting the time dependence of the derivative of the concentration natural logarithm in order of the time natural logarithm shown in Figure~\ref{e_zero} \cite{Goncalves04:a}.
  We plot the derivative versus time for all $p$ values for the zero bias, isotropic diffusion case shown in Fig.~\ref{e_zero}.
  For $t<1$, the various derivatives are not sensitive to the value of $p$, only spreading slightly at the end of the first stage.
  For $t>1$, the system undergoes a transient stage until $t\simeq10^3$ time units before it reaches the diffusion-limited regime for all values of $p\geq0.65$, while for $p=p_c$, $0.61$ the transient time is $\simeq50$.
  The differences in the kinetics for different values of $p$ stems from the fact that the substrates geometrical properties change from fractal, scale-invariant behavior at $p_c$ to a translational invariant\cite{Bunde91}, and also from the domain formation due to local density fluctuations in particle number.
  Time spent in the transient stages may last long enough to prevent a significantly extended diffusion-limited regime from being observed, even for a lattice of size $1024^2$ sites, before late stages kinetics kicks in, which shows difficulties simulating these systems.

  Regarding the strong bias case, the differences relative to the isotropic case emerge right after the first stage as shown in Fig.~\ref{c_high} for all values of $p$, but the compact system, where the overall behavior follows the unbiased diffusion case.
  In contrast to the isotropic case, the transition from the transient to the diffusion-limited regime is now more incisive, and the transition times also vary with $p$.
  Therefore, the effect of a strong bias is to enhance the effect of topological restrictions of the substrate on the kinetics.
  Such an effect of the bias is to slow down the kinetics, a counter intuitive behavior, since the presence of a bias generates a net drift of particles along its direction.
  As can be seen from Fig~\ref{e_high}, the curve for $p=0.8$ shows faster decay than the isotropic case over the transient regime.
  These differences are confirmed by the presence of a well defined peak in the transient stage, as shown in Fig.~\ref{e_high}, followed by \emph{flat terraces}.
  The height and width of the peak of time derivative, see Fig.~\ref{e_high}, now also depends on substrate occupancy.
  For substrate occupancies, values of $p\simeq1$, substrate imperfections are comparable to obstacles.
  However, for occupancies values of $p$ near $p_c$, the substrate is constituted by topological \emph{traps} with few or no exits in the direction of the bias, so particles take a long time to escape from such \emph{traps}.
  Moreover, in the diffusion-limited regime, the exponent rapidly converges to a value, different from its zero bias counterpart for all values of $p$, but $p=1$ \footnote{More extended accounts of this line of reasoning and of the exponent values in the thermodynamic limit will be published in a different article, due to limitations in article size.}.

  We finalize the presentation of our results by presenting the concentration time evolution for various values of the applied bias for a substrate occupancy of $0.65$, as shown in Figure~\ref{c_65}.
  We observe with increasing values of $\eta$, i.e., of the bias, a clear slowing down of the kinetics.
  The slow down of the kinetics is so accentuated that simulations cannot last long enough to observe late stage kinetics.
  In Fig.~\ref{e_65}, we show the time dependence of the numerical derivative, as defined above, for various values of $\eta$, namely, $0$, $0.4$, $0.8$, and $0.9$.
  Notice, the wide variation of late stage kinetic exponents, keeping in mind that for values of $\eta>0$, one can only make rough estimates, since we did not enter deep enough into the late stage kinetics.
  Finally, we present in Table~\ref{alpha}, a summary of the values of the exponent $\alpha$.

 \newpage
 \section{Conclusion}\label{conclusion}
  Though non-compact substrates are quite common in nature, little effort has been put on the characterization of effects of topological restrictions on the kinetics.
  We performed a series of simulations on a square lattice with $1024^2$ sites for various values of substrate occupancy ranging from the percolation threshold to a compact system.
  Our preliminary results show non-trivial substrate occupancy effects on the various stages of the kinetics.
  Further inclusion of bias, i.e., preferred diffusion along one the directions of a driving field, enhances the topological constraints present in the substrate.
  One observes substantial slow down of the kinetics for values of the bias, $\eta$, above $0.4$.
  Substrate topological restrictions, together with the presence of a bias, enhance early times reactivity, but slows it down at long times.
  Our results can be of relevance to researchers working on heterogenous reactions on surfaces and more generally to researchers working on non-compact substrates problems.

 \begin{acknowledgments}
  This research has been partially supported by PRODEP~III (Application 1/5.2/PRODEP/1996 University Advanced Education - Student~10~(NJG)) and Funda\c{c}\~{a}o para a Ci\^{e}ncia e a Tecnologia (grant POCTI/CTM/41574/2001 - Computational Nanophysics).
 \end{acknowledgments}

 %GATHER{E-MRS2004.bib}   % For Gather Purpose Only
 %GATHER{E-MRS2004.bbl}   % For Gather Purpose Only
 \bibliography{E-MRS2004}
 \newpage
 \begin{figure}[p]
  \caption{
   Rules of the $A+B->\emptyset$ diffusion-limited reaction with same species excluded volume interaction on fractal percolating clusters.
    a) Typical configuration at early times.
       Dark grey regions represent the percolating cluster, light grey regions represent remaining clusters, and white regions represent empty sites.
       The image represents a cutout of a larger system of $1024^2$ sites.
    b) From the local configuration, highlighted in a), there is a $B$-particle for which a hopping attempt can lead to:
       i)   a diffusion - upwards;
       ii)  an excluded volume - rightwards;
       iii) bouncing off from the boundary - downwards;
       iv)  a reaction - leftwards.
            In all cases i)-iv) light grey squares represent next-nearest neighbors, which do not directly contribute to the kinetics.
   \label{model}
  }
 \end{figure}
 \begin{figure}[p]
  \caption{
   Single-species concentration versus time log-log plot at zero bias for various substrate occupancies, $p$, on a $1024^2$ sites system.
   \label{c_zero}
  }
 \end{figure}
 \begin{figure}[p]
  \caption{
   Representation of the local slopes of $-\mbox{d}(\log c(t))/\mbox{d}(\log t)$, with $\eta = 0$, for various substrate occupancies.
   \label{e_zero}
  }
 \end{figure}
  \begin{figure}[p]
  \caption{
   Single-species concentration versus time log-log plot at a bias of $0.9$ for various substrate occupancies, $p$, on a $1024^2$ sites system.
   \label{c_high}
  }
 \end{figure}
 \begin{figure}[p]
  \caption{
   Representation of the local slopes of $-\mbox{d}(\log c(t))/\mbox{d}(\log t)$, with $\eta = 0.9$, for various substrate occupancies.
   \label{e_high}
  }
 \end{figure}
 \begin{figure}[p]
  \caption{
   Single-species concentration versus time log-log plot at zero bias for a substrate occupancies of $0.65$, on a $1024^2$ sites system.
   \label{c_65}
  }
 \end{figure}
 \begin{figure}[p]
  \caption{
   Representation of the local slopes of $-\mbox{d}(\log c(t))/\mbox{d}(\log t)$, for various values of $\eta$, and for a substrate occupancy of $0.65$.
   \label{e_65}
  }
 \end{figure}

 \newpage
 \begin{table*}[h]
  \caption{Exponent $\alpha$ in the diffusion-limited regime for each simulated pair of substrate occupancy, $p$, and anisotropy, $\eta$.}
  \label{alpha}
  \begin{ruledtabular}
   \begin{tabular}{lcccc}
    & \multicolumn{4}{c}{$\eta$} \\
    $p$ & $0.0$ & $0.4$ & $0.8$ & $0.9$\\ \hline
    $p_c$ & $0.36\:\pm\:0.06$ & $0.07\:\pm\:0.02$ & $0.06\:\pm\:0.01$ & $0.05\:\pm\:0.01$ \\
    $0.61$ & $0.48\:\pm\:0.04$ & $0.10\:\pm\:0.01$ & $0.06\:\pm\:0.01$ & $0.06\:\pm\:0.01$ \\
    $0.63$ & $0.51\:\pm\:0.05$ & $0.16\:\pm\:0.02$ & $0.08\:\pm\:0.01$ & $0.07\:\pm\:0.01$ \\
    $0.65$ & $0.55\:\pm\:0.05$ & $0.24\:\pm\:0.03$ & $0.11\:\pm\:0.01$ & $0.09\:\pm\:0.01$ \\
    $0.70$ & $0.52\:\pm\:0.04$ & $0.58\:\pm\:0.04$ & $0.17\:\pm\:0.04$ & $0.17\:\pm\:0.02$ \\
    $0.80$ & $0.55\:\pm\:0.04$ & $0.64\:\pm\:0.07$ & $0.50\:\pm\:0.17$ & $0.27\:\pm\:0.10$ \\
    $1$ & $0.52\:\pm\:0.07$ & $0.57\:\pm\:0.04$ & $0.58\:\pm\:0.04$ & $0.54\:\pm\:0.03$ \\
   \end{tabular}
  \end{ruledtabular}
 \end{table*}

 \begin{figure}[p]
  \vspace{7cm}
  \includegraphics[width=16cm]{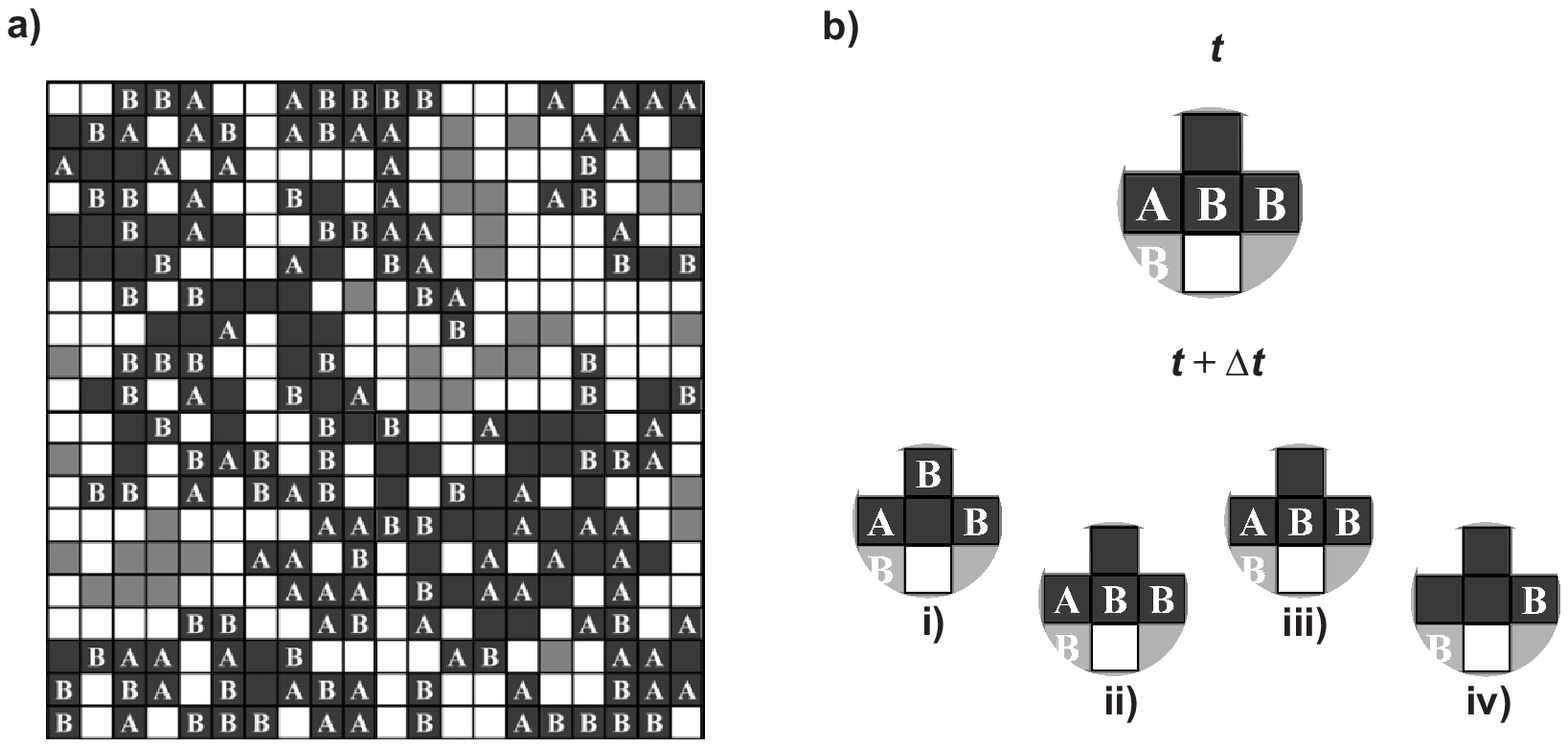} \\
  \vspace{8cm}
  \begin{flushleft}
   {\bf\Huge Figure~\ref{model}}
  \end{flushleft}
 \end{figure}

 \newpage
 \begin{figure}[p]
  \vspace{5cm}
  \includegraphics[width=16cm]{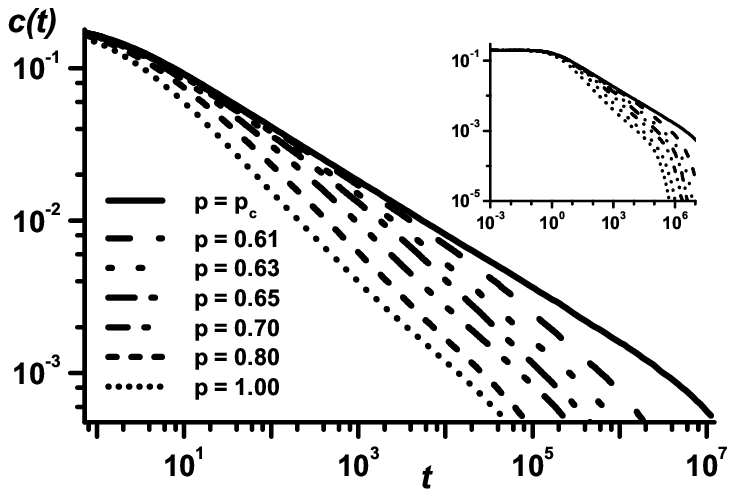} \\
  \vspace{5cm}
  \begin{flushleft}
   {\bf\Huge Figure~\ref{c_zero}}
   \end{flushleft}
  \end{figure}

 \newpage
 \begin{figure}[p]
  \vspace{5cm}
  \includegraphics[width=16cm]{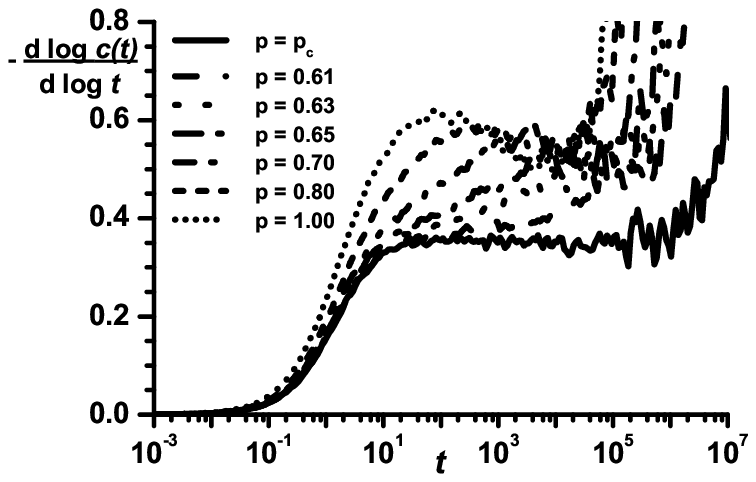} \\
  \vspace{5cm}
  \begin{flushleft}
   {\bf\Huge Figure~\ref{e_zero}}
  \end{flushleft}
 \end{figure}

 \newpage
 \begin{figure}[p]
  \vspace{5cm}
  \includegraphics[width=16cm]{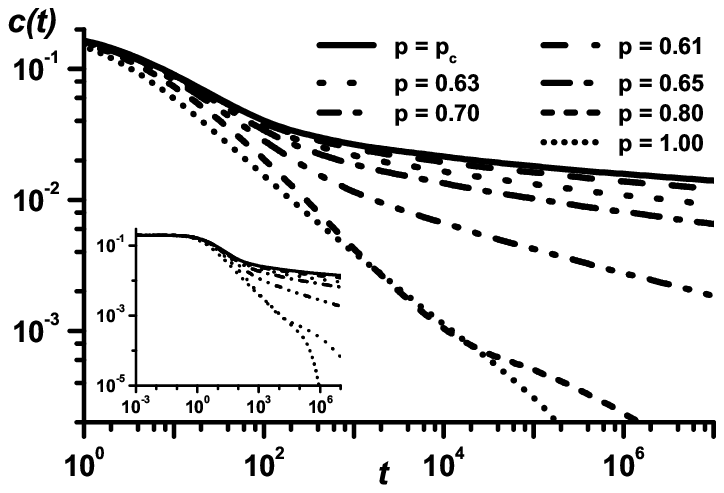} \\
  \vspace{5cm}
  \begin{flushleft}
   {\bf\Huge Figure~\ref{c_high}}
  \end{flushleft}
 \end{figure}

 \newpage
 \begin{figure}[p]
  \vspace{5cm}
  \includegraphics[width=16cm]{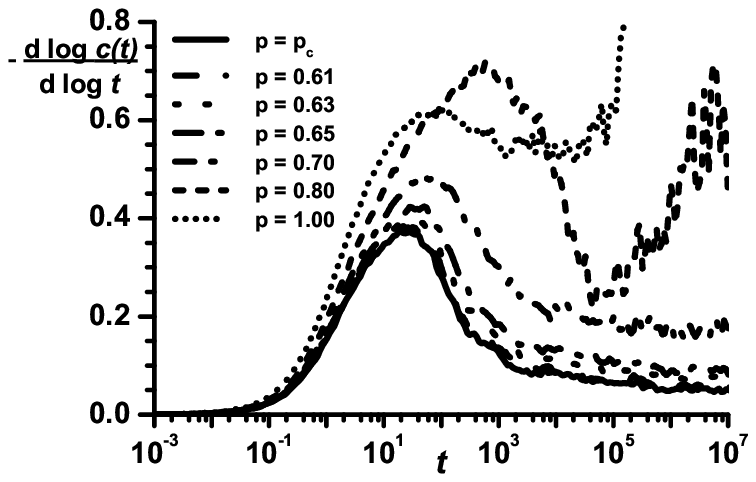} \\
  \vspace{5cm}
  \begin{flushleft}
   {\bf\Huge Figure~\ref{e_high}}
  \end{flushleft}
 \end{figure}

 \newpage
 \begin{figure}[p]
  \vspace{5cm}
  \includegraphics[width=16cm]{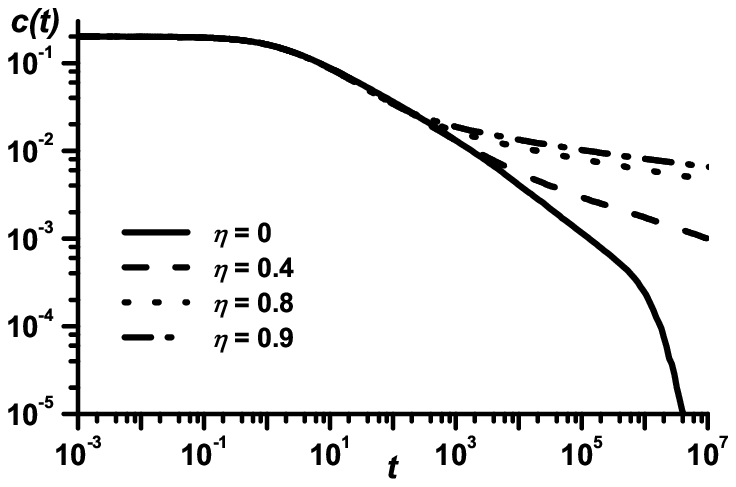} \\
  \vspace{5cm}
  \begin{flushleft}
   {\bf\Huge Figure~\ref{c_65}}
  \end{flushleft}
 \end{figure}

 \newpage
 \begin{figure}[p]
  \vspace{5cm}
  \includegraphics[width=16cm]{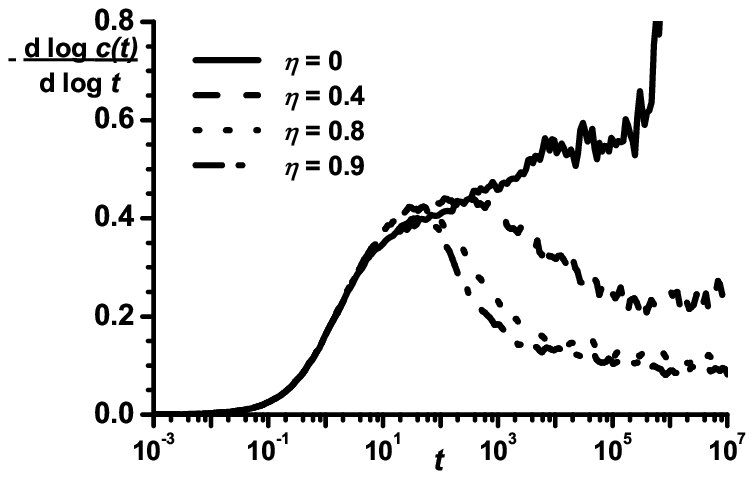} \\
  \vspace{5cm}
  \begin{flushleft}
   {\bf\Huge Figure~\ref{e_65}}
  \end{flushleft}
 \end{figure}

\end{document}